# Lifshitz Transition and Band Structure Evolution in Alkali Metal Intercalated 1T'-MoTe$_2$

Joohyung Park,[1] Ayan N. Batyrkhanov,[1] Jonas Brandhoff,[3] Felix Otto,[3] Marco Gruenewald,[3] Maximilian Schaal,[3] Saban Hus,[4] Torsten Fritz,[1,3] Florian Göltl,[2] An-Ping Li[4], and Oliver L.A. Monti[1,5,*]

**Affiliations**

[1] Department of Chemistry and Biochemistry, University of Arizona, Tucson, Arizona, 85721, USA

[2] Department of Biosystems Engineering, University of Arizona, Tucson, Arizona, 85721, USA

[3] Institute of Solid State Physics, Friedrich Schiller University Jena, 07743 Jena, Germany

[4] Center for Nanophase Materials Sciences, Oak Ridge National Laboratory, Oak Ridge, Tennessee 37831, USA

[5] Department of Physics, University of Arizona, Tucson, Arizona, 85721, USA

*Email: monti@arizona.edu

## Abstract

In van der Waals materials, coupling between adjacent layers is weak, and consequently interlayer interactions are weakly screened. This opens the possibility to profoundly modify the electronic structure, e.g., by applying electric fields or with adsorbates. Here, we show for the case of the topologically trivial semimetal 1T'-MoTe$_2$ that potassium dosing at room temperature significantly transforms its band structure. With a combination of angle-resolved photoemission spectroscopy, scanning tunneling microscopy, x-ray photoemission spectroscopy, and density functional theory we show that i) for small concentrations of K, 1T'-MoTe$_2$ undergoes a Lifshitz transition with the electronic structure shifting rigidly, and ii) for larger K concentrations 1T'-MoTe$_2$ undergoes significant band structure transformation. Our results demonstrate that the origin of this electronic structure change stems from alkali metal intercalation.

# Introduction

Modifying and controlling the properties of quantum materials by external stimuli or by interfacing them with other materials offers enormous opportunities to create materials by design for novel functionalities with electronic and spintronic applications. This is particularly true for van der Waals layered materials where such control is available to an unsurpassed extent, with the aim of tailoring electronic correlations. To achieve this goal, many different approaches have been employed: For atomically thin materials, strain may be used to control, e.g., the bandgap [1,2]; doping may lead to significant electronic structure changes [3,4]; formation of layered heterostructures with other 2D materials can reveal novel exotic electronic phases [5-7]; proximitization with organic molecules may manipulate spin and other degrees of freedom [8,9]; and ultrafast laser excitation may access hidden phases [10,11]. A particularly simple and effective approach is the modification by adsorption of alkali metal atoms, whose low ionization energy enables strong electron doping, and whose small size may allow for intercalation and strong hybridization within the van der Waals gap [12-16]. The consequences of modifying van der Waals materials with alkali atoms vary enormously, depending significantly on the specifics of the pristine phase, and range from simple doping to drastic changes in the band structure.

Though alkali metal deposition has long been used in many studies, new insights keep appearing. Whilst the case of van der Waals layered semiconductors has been studied extensively, the situation for semimetals remains less clear. Indeed, recent evidence suggests that alkali metals can act as electron dopants and chemical gates in van der Waals layered semimetals [12,13,17]. This is important because the charge carrier density in semimetals can control associated many-body phenomena such as superconductivity, charge-density waves, and quantum hall states [18-20]. In the case of the semimetal $T_d$-$WTe_2$, it was shown that alkali dosing not only increases the electron concentration but also that intercalation induces shear displacement to change the crystal symmetry [21]. In contrast, surface doping of potassium at low temperatures on $T_d$-MoTe$_2$ results in quantum confinement, electronically separating the topmost layer MoTe$_2$ from the bulk. [22]. More generally, the lack of a fundamental gap and the increased charge carrier screening in semimetals together with a typically near-balanced coexistence of electron and hole pockets means that the effects of alkali adsorption and potentially intercalation remain intriguing. 1T'-MoTe$_2$, the room temperature phase of $T_d$-MoTe$_2$ [23, 24], is ideally poised to investigate this question further, in particular since alkali diffusion and the associated thermodynamics may lead to competition between these two polytypes.

Further, the topology of the Fermi surface may change rather abruptly, marked by the sudden appearance or disappearance of disconnected pockets. This phenomenon, which is not accompanied by a change in the crystal symmetry, is called a Lifshitz transition, and can be the result of pressure, electron concentration, strain or temperature. Such a transition leads to abrupt changes in thermodynamic, elastic, and transport properties [25] as well as additional many-body phenomena [26, 27]. Methods that map the Fermi surface such as angle-resolved photoemission spectroscopy (ARPES) or Shubnikov-de Haas Oscillations [28] are particularly suited to demonstrate Lifshitz transitions.

Here, we use ARPES and ultraviolet photoemission spectroscopy (UPS) to show that room-temperature potassiation of 1T'-MoTe$_2$ immediately gives rise to a Lifshitz transition at the $\bar{Y}$ point. We also show that the band structure evolution upon potassiation proceeds in two stages: First, a rigid band shift at lower potassium levels, followed by a significant transformation at higher potassium levels. By combining density functional theory (DFT), angle-resolved x-ray photoemission spectroscopy (AR-XPS), and scanning tunneling microscopy (STM), we argue that these changes are caused primarily by potassium intercalation, ruling out other proposed explanations. This study demonstrates the potential of alkali atoms to tailor electronic properties even in layered semimetals, creating new materials based on physics that go far beyond mere doping.

The manuscript is structured as follows: First, we present ARPES data showing how the electronic structure of 1T'-MoTe$_2$ changes upon potassiation. Next, we computationally study the thermodynamics and structural changes of 1T'-MoTe$_2$ as a result of potassiation and use these insights together with ARPES and other surface-sensitive techniques to demonstrate the connection between band structure evolution and potassium deposition. This allows us to draw a comprehensive picture of the impact of potassiation on 1T'-MoTe$_2$ and discuss the implications of our results.

## Methods

### Sample Preparation

The 1T'-MoTe$_2$ crystal was sourced from HQ Graphene and mounted to an Omicron style sample flag using conductive double-sided copper tape. The crystal was freshly exfoliated under ultrahigh vacuum (~ $10^{-9}$ mbar) using the Scotch tape method, and then introduced to better vacuum (~ $10^{-10}$ mbar). Potassium was deposited at ambient temperature from a SAES getter source in a separate preparation chamber with a typical base pressure of $1 \times 10^{-10}$ mbar. Stepwise deposition cycles were taken to increase the amount of potassium, followed immediately by acquisition of the ARPES maps to minimize the impact of potassium diffusion into the bulk crystal. In this way, spectra remained stable over the course of successive sequential potassiation steps.

### LEED Image Acquisitions and Analysis

Low-energy electron diffraction (LEED) images were acquired using a SPECTALEED (Omicron) at room temperature. Post acquisition image correction such as distortion correction was done by the freely available software LEEDCal [29].

### DFT Computations

All calculations were carried out using the Vienna Ab-Initio Simulation package (VASP) [30, 31], a plane wave code using PAW pseudopotentials [32, 33], with the PBE functional [34] and Tkatchenko-Scheffler dispersion corrections [35], and spin-orbit coupling is not considered. In our calculations we used a $k$-point grid of $15 \times 9 \times 3$ for a single unit cell. In slab calculations, $k$-points in the $z$-direction were reduced to 1. In structural optimizations, the $x$ and $y$ lattice vectors in a $1 \times 2$ unit cell were optimized manually first, then the $z$-direction was optimized manually as well. As a final optimization step, the unit cell was allowed to relax. Electronic band structures were separately calculated for 40 $k$-points along the $\Gamma Y$, $\Gamma Y$, and $\Gamma S$ direction. For slab calculations, we use the surface-projected high symmetry direction notations $\overline{\Gamma X}$, $\overline{\Gamma Y}$, and $\overline{\Gamma S}$.

## Ultraviolet Photoemission Spectroscopy and Angle-Resolved Photoemission Spectroscopy

The crystal was aligned along the high symmetry directions ($\overline{\Gamma X}$, $\overline{\Gamma Y}$, and $\overline{\Gamma S}$) using LEED prior to the acquisition of each of the ARPES maps. Ultraviolet photoemission spectroscopy (UPS) and angle-resolved photoemission spectroscopy data was collected with a non-monochromatized He discharge lamp (SPECS 10/35, He I$\alpha$) mounted at a 30º angle of incidence in a VG ESCALab MK II photoemission spectrometer, with an instrumental energy resolution of 120 meV unless mentioned separately. All spectra were acquired at room temperature by successively tilting the sample along the polar angle. Due to the large spot size of the He source (~5 mm) and the irregular shape (5 mm × 11 mm) of the MoTe$_2$ crystal, photoemission intensities may vary along different directions. A sample bias of -5 V was applied during the data acquisition. The acceptance angle was ± 1.5º ($\Delta k_\parallel \approx \pm 0.055$ Å$^{-1}$ at the Fermi level). All ARPES maps are presented after integral background subtraction.

## Angle-Resolved X-ray Photoemission Spectroscopy (AR-XPS)

AR-XPS was conducted using a SPECS Surface Nano Analysis system with a PHOIBOS 150 hemispherical analyzer and a three-dimensional delay-line detector. A Focus 500 (SPECS, monochromatized Al K α emission: 1486.71 eV) was used as an x-ray source with a relative angle to the analyzer of 55º. The acceptance angle of the analyzer was ±8º. The amount of potassium dosing for the XPS experiment was determined by a quartz crystal microbalance (Tectra), calibrated to be approx. 40% of the appearance of the (2 × 2) superstructure of K-intercalated 1 ML epitaxial graphene on SiC(0001)[36].

## Scanning Tunneling Microscopy

STM measurements were performed at Oak Ridge National Laboratory with a variable-temperature STM system (Scienta Omicron VT-STM) operated at room temperature with base pressure of $5\times10^{11}$ mbar. Electrochemically etched tungsten tips were used as STM probes. The quality and sharpness of the STM tips were tested with STM/STS measurements on a clean Cu(100) surface before being used on the 1T'-MoTe$_2$ surface.

# Results

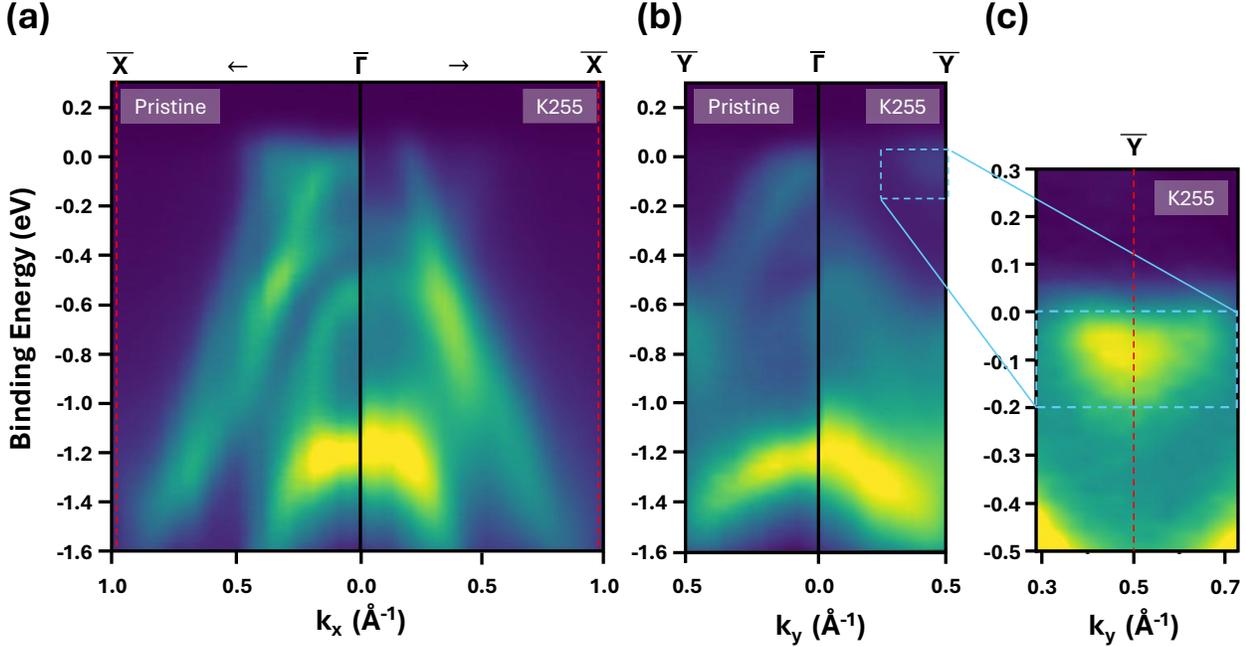

Figure 1. ARPES maps of band structure evolution after potassiation, showing profound band structure changes beyond mere doping. (a) $\overline{\Gamma X}$ map of pristine (left) and heavily K-dosed MoTe$_2$ (K255, right). (b) $\overline{\Gamma Y}$ map of the same (pristine and K255). Vertical lines indicate high symmetry points. (c) A zoom-in ARPES map near the $\overline{Y}$ point, showing the appearance of an electron pocket upon potassiation. Instrumental energy resolution for the ARPES map in (c) is 70 meV.

We first focus on spectroscopy results of the effect of potassium dosing on 1T'-MoTe$_2$. Figure 1 shows ARPES maps that represent the key changes in the band structure. Since K atoms may adsorb at the surface or intercalate, we prefer to label each sample in terms of the K deposition time in seconds rather than in some likely fictitious fractional monolayer coverage: E.g., K255 indicates 255 seconds of cumulative potassiation of the sample. Figures 1 (a) and (b) show the band structure along $\overline{\Gamma X}$ and $\overline{\Gamma Y}$ of pristine 1T'-MoTe$_2$ (left) and heavily potassiated MoTe$_2$ (right). Our ARPES map of pristine 1T'-MoTe$_2$ matches well with previous ARPES data, consistent with the expected high degree of crystalline order observed in LEED (Supporting Information, Figure S1) [24, 37]. Comparing the ARPES map of the K255 structure to the pristine sample, abrupt discontinuities at $\overline{\Gamma}$ suggest that the band structure change is not just a simple rigid band shift. Moreover, the presence of strong spectral weight near the Fermi level at the $\bar{Y}$ point in K255 indicates the emergence of a new electron pocket, clearly evident in Figure 1 (c), changing the connectivity of the Fermi surface.

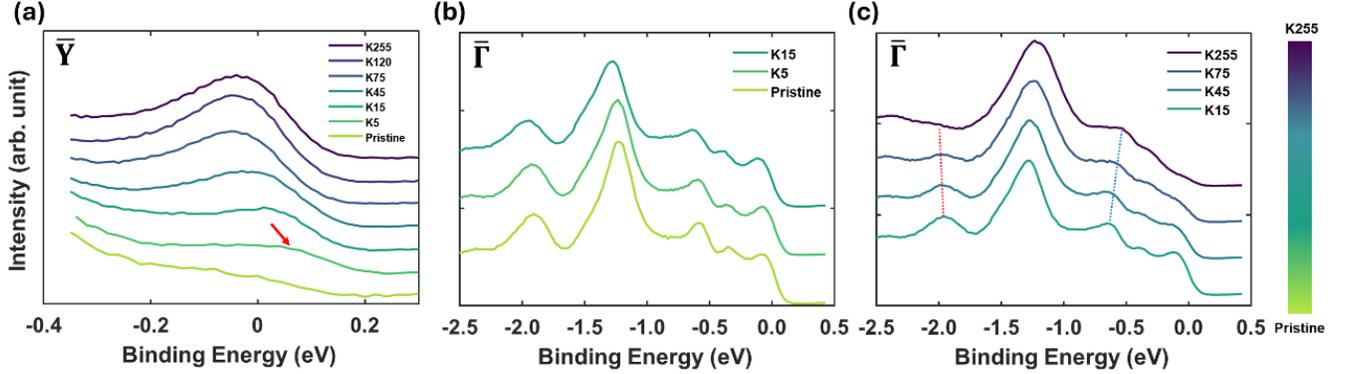

Figure 2. Detailed band structure evolution near $\bar{Y}$ and $\bar{\Gamma}$ with progressions of potassiation in 1T'-MoTe$_2$, (a): showing the emergence of a new electron pocket at $\bar{Y}$. The red arrow indicates the signature of the new electron pocket. The complex changes to the band structure upon potassiation, shown in (b) and (c) at $\bar{\Gamma}$. (b) Small dosing limit: Pristine, K5, and K15. (c) Medium and high dosing limit: K15, K45, K75, and K255 at $\bar{\Gamma}$. Two dashed lines in red and blue indicate complicated band structure evolution. K15 is included in (c) to connect to panel (b). Note that the same colors correspond to the same potassium dosing across different panels (see color bar).

To gain deeper understanding of the electronic structure change, we directly compare spectra with different amounts of potassium for two different points in the Brillouin zone, namely $\bar{\Gamma}$ and $\bar{Y}$. At $\bar{Y}$ (Figure 2 (a)), already the smallest amount of potassium (5s, sample K5) leads to the appearance of spectral weight near $E_F$, and the emergence of a new electron pocket (marked by a red arrow). The intensity of this new band keeps growing until K75 and saturates thereafter. Figures 2 (b) and (c) show the band structure evolution at $\bar{\Gamma}$ in the first 2.5 eV below $E_F$, for the case of low potassium dosing (K5 and K15) and across a wider range of potassium dosing (K15 to K255), respectively. Initially, for lower potassium dosing, the bands shift monotonically to higher binding energies for all the observed features in the spectra. From the fact that the peak shifts are monotonic and that the spectra show only moderate broadening, the low concentration potassium limit can be understood as doping, though, accompanied by the growth of a new electron pocket at $\bar{Y}$. More drastic changes occur at higher doses of K on 1T'-

MoTe$_2$ as shown in Figure 2 (c): *i)* The peak positions shift non-uniformly, e.g., the band at a binding energy of -1.9 eV shifts further below $E_F$ (red dashed line), whereas the band at a binding energy of -0.6 eV shifts closer to $E_F$ (blue dashed line). *ii)* Peaks broaden with increasing K coverage. *iii)* The photoemission intensity of the bands near the Fermi level diminishes at the $\bar{\Gamma}$ point. As we will show below from STM data and in accordance with our DFT results, these changes are not due to surface degradation caused by disordered adsorption of K, nor the growth of an ordered K superstructure. Instead, we attribute these changes in the band structure to changes in the electronic structure and the photoemission matrix element.

Overall, changes of the band structure occur throughout the Brillouin zone, including a continued growth of the electron pocket at $\bar{Y}$ (Figure 2(a)). This complex band evolution is likely a remarkable consequence of the weak van der Waals interlayer interaction in 1T'-MoTe$_2$ and the outsized impact of strong electron donors such as K even in a semimetal such as MoTe$_2$. It is likely a feature of other layered semimetals as well, although there the details may vary depending on the materials and preparations [12, 13, 21].

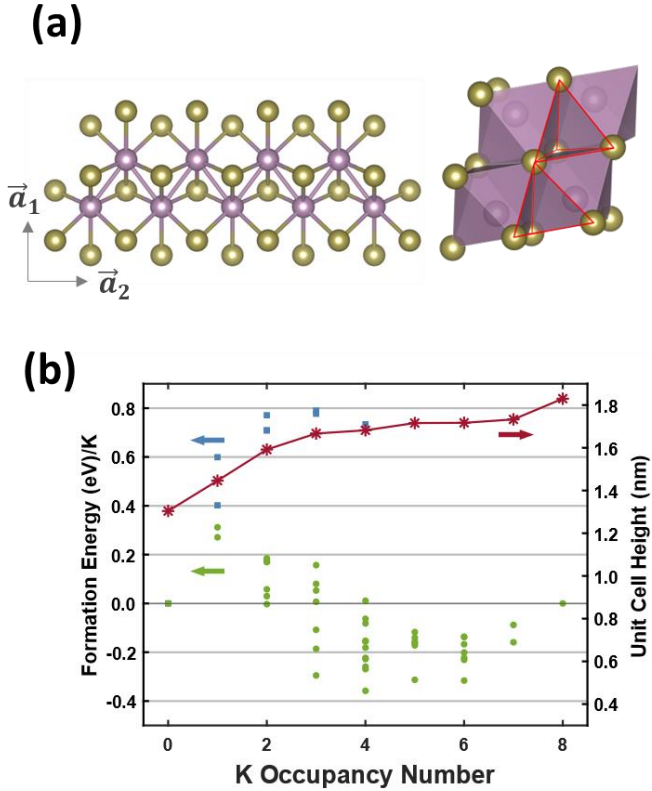

Figure 3. Structural considerations and DFT analysis showing potassiation sites, swelling of the MoTe$_2$ crystal upon potassiation, and the energetic competition of intercalation vs. surface adsorption. (a) Left: Top-down view of the crystal structure of 1T'-MoTe$_2$. Mo: light purple, Te: light gold. Right: Cartoon of the crystal structure. Red triangles show potassiation sites. (b) DFT calculation of the thermodynamics of potassiation ($U = 0\ eV$). Light green: Formation energy of potassium intercalation as a function of the K occupancy number in a $1 \times 2$ superstructure. Light blue: The same for a surface adsorbate. The scattered dots indicate the formation energy from different configurations in the supercell. Dark red line: Unit cell height per K occupancy number in the optimized bulk structure.

To clarify the mechanism underlying these changes, we computationally investigate the changes to 1T'-MoTe$_2$ upon K deposition using DFT+$U$. We find the best agreement between experiment and DFT for the pristine 1T'-MoTe$_2$ with a $U$ of 2.0 eV, and a $c$-axis unit cell spacing of 14.2 Å. Details considering this parameterization and the comparison with experiment are given in Figures S2 and S3 in the supplemental material. Next, we consider the crystal structure of pristine 1T'-MoTe$_2$ with regards to potassium deposition. The bulk 1T'-MoTe$_2$ unit cell consists of two MoTe$_2$ sheets, and each MoTe$_2$ sheet contains two inequivalent Mo atoms that establish a zig-zag network in the $\vec{a}_2$ direction (Figure 3 (a)). Each Mo atom is coordinated by six Te atoms forming a distorted octahedron of which three Te atoms create the triangular face of the surface 2D sheet (light purple faces in the right hand side of cartoon Figure 3 (a)). Adjacent to each octahedron is a tetragonal void, of which there are two in each unit cell (highlighted with red lines in Figure 3 (a)). We hypothesize that the voids can take potassium atoms, viz. we assume a maximum K loading of 1 K / Mo in our calculations. The relative spacing of the layers is determined by the *c*-axis of the unit cell, normal to the MoTe$_2$ layers.

We then seek to understand the nature of the thermodynamically most favored structure of K-MoTe$_2$. Here, we consider two situations as relevant limiting cases, namely 1) surface adsorption and 2) bulk intercalation. In order to study different K loadings and search within a large structural space, we initially focus on a $1 \times 2$ supercell to allow nearest neighboring lattice interactions along the molybdenum zig-zag network. For surface adsorption, this leads to a maximum of four K atoms per supercell, and for bulk intercalation to a maximum of eight K atoms per supercell. We then investigate all possible permutations for one to four (or eight) K atoms per supercell and show the formation energy / K-atom as well as the *c*-axis value of the intercalated bulk unit cell in Figure 3(b). We define the formation energy per potassium as

$$E_{form/K} = (E_{MoTe_2+nK} - E_{MoTe_2} - n \times \mu_K^{MoTe_2+8K})/n \qquad (1)$$

with

$$\mu_K^{MoTe_2+8K} = (E_{MoTe_2+8K} - E_{MoTe_2})/8 \qquad (2)$$

for the example of potassium intercalated bulk 1T'-MoTe$_2$. Here *n* is the occupancy number, $E_{MoTe_2+nK}$ indicates the total energy of the MoTe$_2$ lattice with a certain potassium occupancy number *n*, and $\mu_K^{MoTe_2+8K}$ is the chemical potential of potassiated 1T'-MoTe$_2$ with an occupancy number of eight, as defined in eq. (2). This sets the formation energy of K to zero for occupancy numbers 0 and 8, thereby defining maximum potassiation as the relevant thermodynamic reference state [38]. $E_{form/K}$

expresses therefore the energy gain per potassium atom (surface adsorption or intercalation) relative to the chemical potential of the fully potassiated bulk 1T'-MoTe$_2$ crystal. Note that in the case of the surface adsorbate, $n_{max} = 4$ instead of 8. Also, these calculations are carried out for $U = 0$ eV, since the value of $U$ has no impact on the overall behavior of the formation energy.

Our DFT results illustrate two important points: First, surface adsorption of K on MoTe$_2$ is *always* less favorable than intercalation. Second, an occupancy number of around four in the intercalation scenario is thermodynamically the most favorable configuration, though three or higher cannot be excluded, as they all offer thermodynamic gain. In contrast, introducing only one or two K atoms in the system involves either an energy penalty or no gain in stability. This implies that at low K loadings, K atoms may not spread uniformly into the lattice but rather form local intercalated clusters of higher potassium density. This local clusters picture might qualitatively explain small peak broadenings found in both K5 and K15 samples in Figure 2 (a). Similar findings on thermodynamically lower stabilities at smaller alkali metal loadings have also been reported for other systems, though the formation energy trends are not universal due to sometimes different reference states considered for calculating the formation energy [14, 38].

The calculated crystal structure from the most stable configuration also provides evidence that three to six is the preferred occupancy number: When considering the *c*-axis dimension upon intercalation (see Figure 3 (b)), we find a consistent increase in unit cell height at smaller occupancy numbers. However, as the occupancy number reaches four, the *c*-axis dimension levels off, before increasing again at seven and eight. Structurally, this indicates an effective increase of the layer separation of adjacent MoTe$_2$ layers that saturates around four K, resulting in severe consequences for the electronic structure.

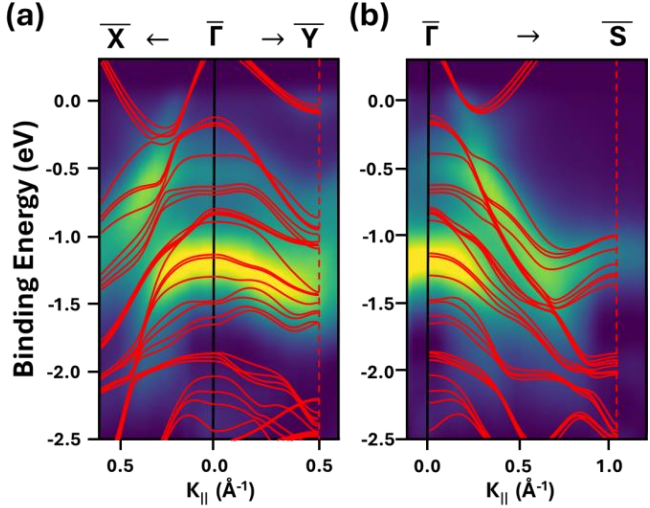

Figure 4. Overlay between the optimized potassium intercalated band structure (occupancy number of four, red lines) and the K255 ARPES maps, showing good agreement between experiment and DFT results. (a) High symmetry directions $\overline{\Gamma X}$ and $\overline{\Gamma Y}$. Note that for clarity we only show the cut in the $\overline{\Gamma X}$ direction. (b) Another high symmetry direction $\overline{\Gamma S}$, diagonal in the Brillouin zone.

With these structural and thermodynamic considerations in mind, we calculate the band structure for the half-filled (occupation number 4), fully filled (occupation number 8), and surface-potassiated 1T'-MoTe$_2$ for comparison with our experimental data. For this purpose, we focus on a four-layer slab with a $1 \times 1$ unit cell (see Figure S3 for the DFT optimized band structures of pristine 1T-MoTe$_2$ using slab calculations). Slab calculations are appropriate since they reflect the surface-sensitive nature of photoemission spectroscopy. Also, the $1 \times 1$ unit cell captures the nature of the most stable K configurations and describes the experimentally observed Brillouin zone size as we do not observe signatures of superlattice formation and zone folding. We note that we neglect spin-orbit coupling, as our focus is on the potassium-induced changes to the band structure. When finding the best matching occupancy number compared to the ARPES map of the K255 sample, we revisit the Hubbard $U$, since there is a possibility that the presence of potassium changes the chemical nature of the Mo atoms and the d-band occupancy. However, we find that a Hubbard $U$ value of 2.0 eV is still optimal. We then compare the DFT band structures to the ARPES maps of the K255 sample. As already expected from the thermodynamic data, we find that the band structures for occupancy number of four (half-filling, 0.5 K / Mo atom) show good agreement with our experiments in all the momentum space directions investigated (see Figure 4 (a) and (b)). Additionally, we note that the possibility of full occupancy (occupation number eight) and surface-adsorbed K as an alternative explanation of the experimentally observed band structures can be safely excluded by comparing, e.g., the bands found near the high symmetry points $\bar{Y}$ and $\bar{S}$, where unacceptable spurious bands appear for those two limits (see Figure S4 (b) – (d) for the relevant band structures).

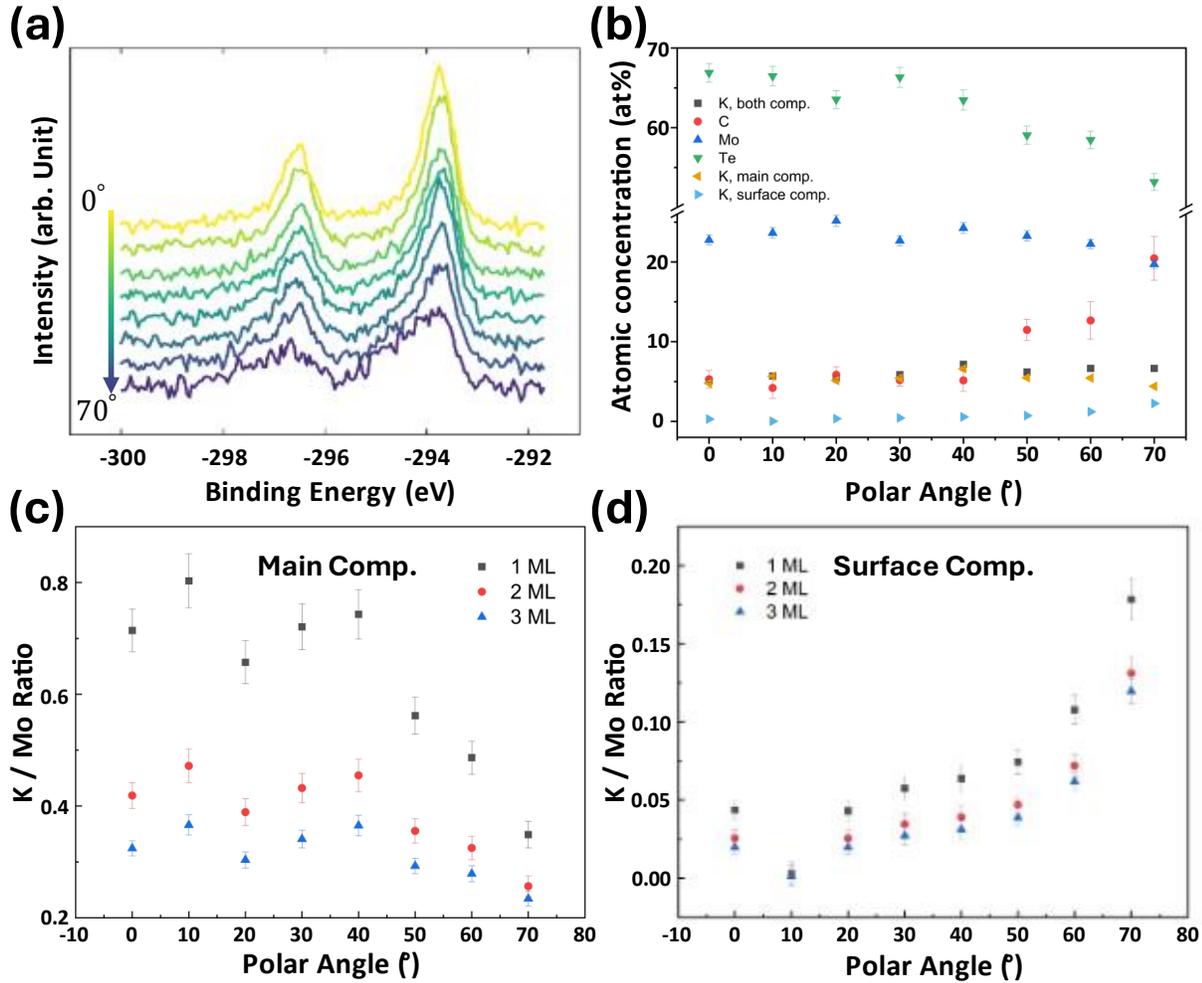

Figure 5. AR-XPS of K-MoTe$_2$ shows the predominantly bulk nature of K and a lesser surface adsorbate component, consistent with intercalation of K. (a) Angle-resolved K 2p spectra of a highly potassiated sample. From top to bottom: Emission angle of 0° to 70°, with 10° step size. (b) Angular variation of the detected elemental atomic concentrations. (c) and (d) K to Mo ratios assuming an MoTe$_2$ thickness of 1 ML, 2 ML, and 3 ML, where the main component K is used in (c), and the surface component K is used in (d).

We use angle-resolved XPS (AR-XPS) to test our hypothesis of predominant K intercalation. By varying the photoemission take-off angle, the XPS data gives depth-related information from near the surface of 1T'-MoTe$_2$. Figure 5(a) shows angle-resolved background-subtracted AR-XPS of K 2p. From the K 2p peaks obtained at different take-off angles, we observe that the 2p$_{3/2}$ and 2p$_{1/2}$ profiles develop an additional component at higher emission angles, in coexistence with the main component. This is suggestive of the existence of two different chemical environments of the K atoms. We propose that the two components correspond to a minor contribution of surface adsorbate (surface component) in the presence of an intercalated species (main component). Using relative sensitivity factors [39], Figure 5(b) shows atomic concentrations determined for each of these two components as well as for other detected elements as a function of the take-off angle. Recognizing that there is a minor surface-confined carbon contamination, neither Te nor Mo features show a pronounced angle dependent peak shape, consistent with the homogeneous bulk nature of $MoTe_2$. For potassium, the main component is present at all angles. In contrast, the minor component remains almost non-existent at lower angles and increases only at higher angles. This indicates that the minor component is indeed likely to correspond to a small contribution of surface adsorbed potassium. To investigate what the most likely depth might be for each of the potassium components, we construct a model by taking the inelastic mean free path and information depth of XPS into account [40,41]. Figure 5(c) and (d) show the K / Mo ratio over emission angles for three different modeled depths: surface layer, topmost two layers, and topmost three layers of the 1T′-MoTe2 crystal. In this model, the depth where the K / Mo ratio value over emission angle changes the least represents the best estimate of the average depth of the deposited potassium. Of course, an increasing (decreasing) K / Mo ratio at higher polar angles shows a K distribution closer to the surface (deeper in the bulk). From this consideration, we see that the majority of the main potassium component resides indeed away from the surface, likely near the 2nd or perhaps even 3rd layer. Further, to test if the ratio of surface-bound and intercalated K changes with the degree of potassiation, we doubled the amount of the deposited potassium. The relative atomic concentrations do not differ (Figure S5), suggesting that once potassiation of 1T'-$MoTe_2$ reaches a point where maximum thermodynamic gain occurs, e.g., an occupancy number of four as suggested by DFT, additional potassium atoms diffuse to deeper layers that are not observed by XPS. AR-XPS therefore supports the notion of a majority intercalation of potassium below the surface, in the presence of a minor component of a surface adsorbate.

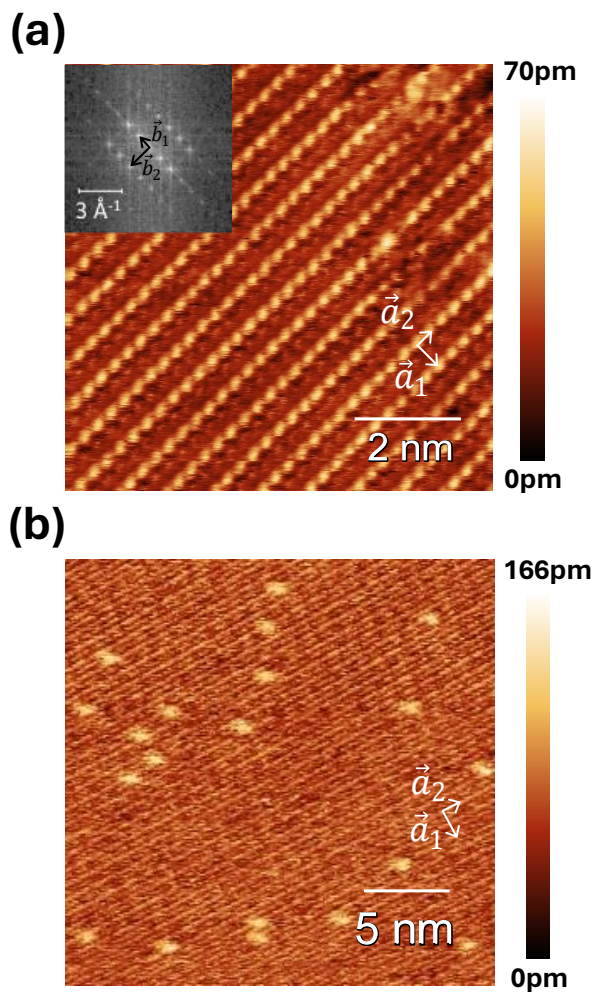

Figure 6. STM image of highly potassiated K-MoTe$_2$ (comparable to K255) at room temperature showing a highly ordered MoTe$_2$ surface with a low defect concentration. (a) Small-area scan with a bias voltage of +100 mV and a current setpoint of 70 pA. The corresponding 2D FFT is shown in the inset. (b) Larger-area scan with a bias voltage of +400 mV and a current setpoint of 10 pA. The unit cell vector lengths are magnified by a factor of 2 in (b).

To further test these findings, we turn to scanning tunneling microscopy (STM). The constant current STM image of the pristine substrate shows highly ordered rows of Te atoms (see Figure S6 (a)). After heavy potassiation, the sample surface exhibits the same Te atomic rows with occasional defects, as shown in Figure 6 (a). The 2D FFT of this and similar topography data reveal a periodic pattern with the calculated reciprocal space vectors of $|\vec{b_1}| = 1.00$ Å$^{-1}$ and $|\vec{b_2}| = 1.94$ Å$^{-1}$, corresponding to lattice vectors of $|\vec{a_1}| = 6.34$ Å and $|\vec{a_2}| = 3.23$ Å, in excellent agreement with the crystal structure of pristine 1T'-MoTe$_2$ [42]. This agrees also with DFT which shows that intercalation does not noticeably change these two dimensions. The defect density is estimated as 1 defect / 120 unit cells from the 100 nm × 100 nm image, indicating the surface remains well preserved, see Figure S7. The STM image in Figure 6 (b) captures the high degree of crystallinity over larger distances, but also manifests the appearance of many bright protrusions (Figure 6 (b)). The change of the tip height along those bright protrusions amounts to an average of 41 ± 9 pm. Bright shallow protrusions of this kind have been assigned as a signature of potassium intercalation in the related WTe$_2$ [43]. We can, however, not exclude that some of these protrusions may rather be native defects also found in pristine T$_d$-MoTe$_2$ [44, 45]. Crucially though, the STM topography does not carry signatures of potassium *surface adsorbates*, expected to lead to much larger height changes [43], nor of potassiation induced significant disorder. This is supported by previous STM studies of alkali atom depositions on 2D materials, where surface adsorbates are only observed at much lower sample temperatures during depositions, likely due to inhibition of intercalation at cryogenic temperatures [13, 26, 46, 47]. Our data thus indicates that the population of surface-adsorbed potassium atoms is indeed very small or non-existent, fully consistent with our AR-XPS, ARPES and DFT results.

## Discussion

In taking together all the experimental data and the DFT calculations, a consistent picture of the consequences of potassiation arises. Potassium atoms prefer intercalation over surface adsorption, swelling the *c*-axis of the crystal, as can be seen from Figure 3 (b). As Figures 1 and 2 demonstrate, this causes significant changes in the band structure. An occupancy number of three or higher is thermodynamically stable, and this is consistent with our findings from the band structure analysis presented in Figure 4, which favors half-filling (occupancy number of four) over surface adsorption or full filling. Excess potassium atoms are driven deeper into the bulk crystal, intercalating between layers farther away from the surface (see, e.g., Figure 5 (b) and Figure S5) to maximize thermodynamic gain. As long as there is room for potassium atoms to diffuse into the bulk, the lattice maintains a partial filling, likely around an occupancy number of four.

The UPS spectrum of K15 at $\bar{Y}$ indicates the appearance of a new electron pocket at $E_F$ and hence a change of the connectivity in the Fermi surface. Such changes are the hallmark of a Lifshitz transition that occurs here already at low potassium dosing, in the limit where most bands shift rigidly. This indicates that the Lifshitz transition is a separate process, apart from the significant band structure evolution observed at higher dosing. This offers opportunities for the Fermiology of K-intercalated 1T'-MoTe$_2$, as changes in the Fermi surface and the associated Lifshitz transition are typically accompanied by a sudden onset of new many-body processes that are not observed in the pristine host material [26, 27]. Indeed, a recent study of monolayer 1T'-MoTe$_2$ on graphene [48] demonstrates the emergence of charge order as a result of a partial charge transfer from graphene to 1T'-MoTe$_2$, where the nesting vector aligns with the $\bar{X}$ axis. In a similar fashion, the Lifshitz transition in bulk 1T'-MoTe$_2$ via potassium intercalation as observed here may also lead to new correlated phenomena.

Alternative sources for the origin of the electron pocket at $\bar{Y}$, different from a Lifshitz transition, should be considered as well. Possible mechanisms include the formation of a free electron band [46], surface potential-induced band bending of the bulk band [13], or the formation of quantum well states as a result of multilayer potassium adsorbate formation [47]. These interpretations are, however, rather unlikely for the present case: First, the electron pocket appears at a high symmetry point away from $\bar{\Gamma}$. The formation of a free electron band or a quantum well away from the center of the Brillouin zone can only occur if other scattering processes are at play [49], and there is no evidence for such scattering. Second, we observe intercalation of potassium atoms rather than significant surface adsorption, rendering a multilayer potassium adsorbate highly unlikely. Third, ultrafast laser excitation has been reported to cause a reduction of the value of Hubbard $U$ term, resulting in a Lifshitz transition similar to the one observed here [10]. In contrast, even for the high potassiation limit we find no necessity to adjust the Hubbard $U$ value. Hence the Lifshitz transition in our case originates from potassiation, suggesting an important role of the interaction between MoTe$_2$ lattice and the intercalated potassium atoms.

Our findings are also important to understand opportunities to access Weyl points at room temperature. Electron transfer from potassium upon intercalation may generate a vertical electric field in the 2D sheets which would break inversion symmetry without the need for inducing a structural shear displacement to the T$_d$ phase. Based on the observed band shifts in our ARPES data, we speculate that it may be possible to shift the Weyl points which are above $E_F$ in T$_d$-MoTe$_2$ to below $E_F$ using intercalation with a strong electron dopant. This may induce a topological phase transition to a new room temperature Weyl phase, at least in the first few potassiated layers. It is, however, known that perturbations to the pristine crystal lattice can change the number of Weyl points or even eliminate them [50-52], and potassium intercalation may do so. Our study suggests that large degrees of potassiation in fact suppress the presence of a hole pocket, which in turn would suppress the type-II Weyl phase. Nevertheless, we believe that proximitization and intercalation with adsorbates has the potential to manipulate electronic and potentially topological phase transitions.

## Conclusions

To conclude, our results show that the different stages of potassiation lead to distinctive changes in the electronic structure of the layered semimetal 1T'-$MoTe_2$. The measured ARPES spectra show the band structure progression from monotonic rigid band shifts to complex band structure evolution. A Lifshitz transition occurs already at low K dosing, independent of the band structure evolution at high potassium dosing. Our DFT study clearly shows that intercalation is preferred over surface adsorption, with an occupancy number near half-filling preferred. The lack of significant surface adsorption and the preference for intercalation is also confirmed by AR-XPS and STM. Using our combined theory and multi-experiment approach, we thus demonstrate that the intercalation of potassium atoms can generate overall substantial band structure changes. Atomic intercalation may thus offer new avenues for tailoring electronic and topological phases.

## Conflict of Interest

The authors declare no competing conflict of interest.

## Acknowledgments

This research is supported by the National Science Foundation under Grant No. NSF CHE-1954571 and by the College of Science of the University of Arizona. Oliver Monti gratefully acknowledges support from the Carl Zeiss Stiftung for a visiting professorship at the Friedrich Schiller Universität Jena. Joohyung Park, Ayan N. Batyrkhanov, and Oliver L. A. Monti appreciate Paul Lee for technical help in the initial potassium deposition setup. This task was carried out at the University of Arizona Department of Chemistry and Biochemistry in the Laboratory for Electron Spectroscopy and Surface Analysis (LESSA) Facility, RRID:SCR_022885. Florian Göltl thanks the College of Agriculture and Life Science at the University of Arizona for their support. We acknowledge computational time at the National Energy Research Scientific Computing Center (NERSC), a DOE Office of Science User Facility supported by the Office of Science of the U.S. Department of Energy, contract DE-AC02-05CH11231. The STM data set was collected at the Center for Nanophase Materials Sciences (CNMS), which is a U.S. Department of Energy, Office of Science User Facility at Oak Ridge National Laboratory.

## For Table of Contents Only

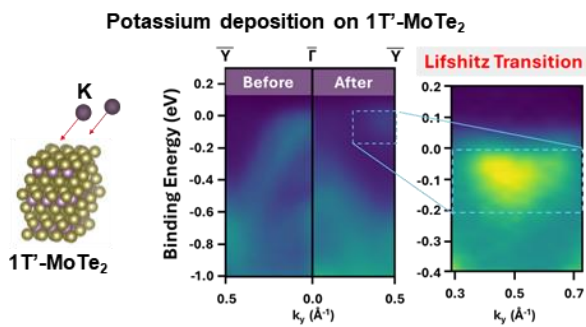

# Supporting Information

# Lifshitz Transition and Band Structure Evolution in Alkali Metal Intercalated 1T'-MoTe$_2$

Joohyung Park,[1] Ayan N. Batyrkhanov,[1] Jonas Brandhoff,[3] Felix Otto,[3] Marco Gruenewald,[3] Maximilian Schaal,[3] Saban Hus,[4] Torsten Fritz,[1,3] Florian Göltl,[2] An-Ping Li,[4] and Oliver L.A. Monti[1,5,*]

**Affiliations**

[1] Department of Chemistry and Biochemistry, University of Arizona, Tucson, Arizona, 85721, USA

[2] Department of Biosystems Engineering, University of Arizona, Tucson, Arizona, 85721, USA

[3] Institute of Solid State Physics, Friedrich Schiller University Jena, 07743 Jena, Germany

[4] Center for Nanophase Materials Sciences, Oak Ridge National Laboratory, Oak Ridge, Tennessee 37831, USA

[5] Department of Physics, University of Arizona, Tucson, Arizona, 85721, USA

*Email: monti@arizona.edu

# Figure S1. LEED Pattern of Pristine 1T'-MoTe₂

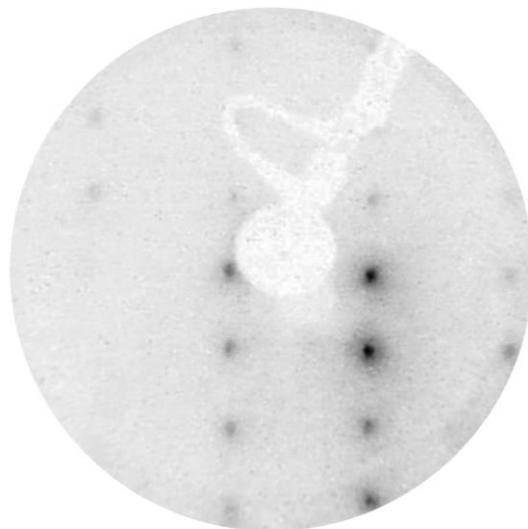

Figure S1. (a) Distortion-corrected LEED pattern of pristine 1T'-MoTe₂ taken with an electron energy of 80 eV and shown in inverted contrast showing the expected lattice for 1T'-MoTe₂. Residual stray magnetic fields dim diffraction spots in the lower left corner and shift the specular point to right.

# Figure S2. DFT Parametrization of Layer Separation and Hubbard $U$ and
# Figure S3. Comparison Between ARPES and Slab Calculation of the Band Structure of Pristine 1T'-MoTe$_2$

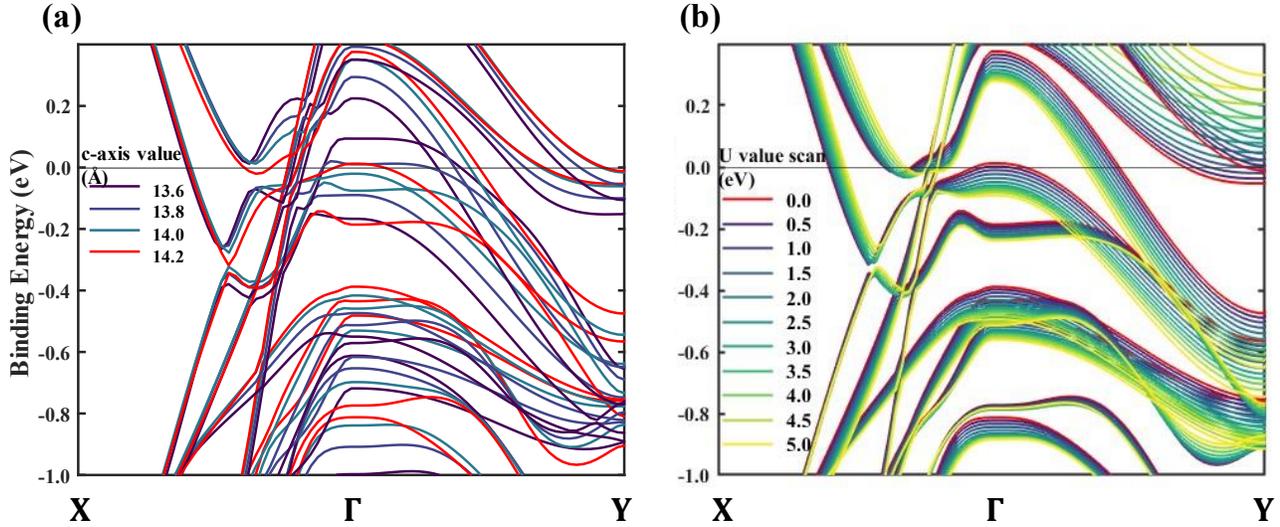

Figure S2. Parametrization of the DFT band structure of pristine 1T'-MoTe$_2$ along the two high symmetry directions $\overline{\Gamma X}$ and $\overline{\Gamma Y}$ by changing (a) the c-axis value (for clarity, only selected structures are shown) with an $U$ value of 0 eV, and (b) changing the $U$ value with the c-axis value of 14.2 Å. The colors correspond to different parameter values. Red colored bands in (a) and (b) share the same c-axis and $U$ values. See text below for the detailed procedure.

One of the major challenges in performing band structure calculations for 1T'-MoTe$_2$ is to correctly describe the energies of the Mo d-states, since the PBE-functional is known to underestimate the gap between these states. One way to address this shortcoming is to artificially increase the separation between the d-states using the $+U$ method. Several different $U$ values for the Mo d-states have been suggested, but no agreement on the correct $U$ value has been reached for 1T'-MoTe$_2$ [1-3]. In addition, the calculated band structures also depend on the layer separation in 1T'-MoTe$_2$. We use a dispersion correction as parameterized by Tkatchenko and Scheffler [4], which overestimates dispersion corrections for the 1T'-MoTe$_2$ system and significantly underestimates the layer separation compared to the experimentally known crystal structure.

To address these shortcomings, we parameterize *(i)* the $U$ parameter and *(ii)* the layer separation in 1T'-MoTe$_2$. These two parameters and their effects are independent of each other. $U$ corrects for a shortcoming in semi-local GGA approximations to correctly describe the d-states of Mo, while the *c*-axis of the unit cell depends on the correct description of long-range dispersion forces, which are by definition not captured correctly by semi-local approximations. By comparing the ARPES maps with the DFT calculations. For this phase in particular, the experimentally observed absence of an electron pocket around the Fermi level at $\bar{Y}$ is an important feature that is not correctly reproduced by pure PBE [3], and we use this as an indicator for the parametrization.

To begin, we fix the $U$ value at 0 eV and vary the c-axis value of the unit cell between 13.0 Å and 14.2 Å. As the resulting bands strongly disperse by this parameter, we only show some select parameters in Figure. S2 (a) for clarity. Here, some bands shift to higher energies, while other bands shift to lower energies, and other bands remain almost unchanged. Focusing on the bands close to the Fermi level at $\bar{Y}$, we find that electron pockets are shifted to higher energies with increased layer separation. The electron pocket size at $\bar{Y}$ shows negligible change as the layer separation reaches 14.0 Å and higher.

We then vary the $U$ parameter for the Mo 3d states between 0.0 eV and 5.0 eV at three layer separation values: 13.8, 14.0, and 14.2 Å. We find that all bands in the relevant energy window are significantly impacted by $U$. Most bands are increasingly shifted to lower energies with $U$, only the unoccupied bands around $\bar{Y}$ are shifted to rise above the Fermi level. Figure S2 (b) shows the case with a layer separation of 14.2 Å. As the $U$ value reaches +2.0 eV, the electron pocket at $\bar{Y}$ lies above the Fermi level, which is a qualifying condition that our ARPES map of the pristine sample requires.

We then switch to a slab calculation method and iterate this procedure, adjusting $U$ and the *c*-axis value (see Figure 4 and related main text for details, see slab calculations results shown in Figure S3) until an optimal match with the experimental ARPES band structure is achieved. Overall, we find that again $U$ of +2.0 eV and layer separation of 14.2 Å offer the best agreement with the ARPES map of pristine 1T'-MoTe$_2$.

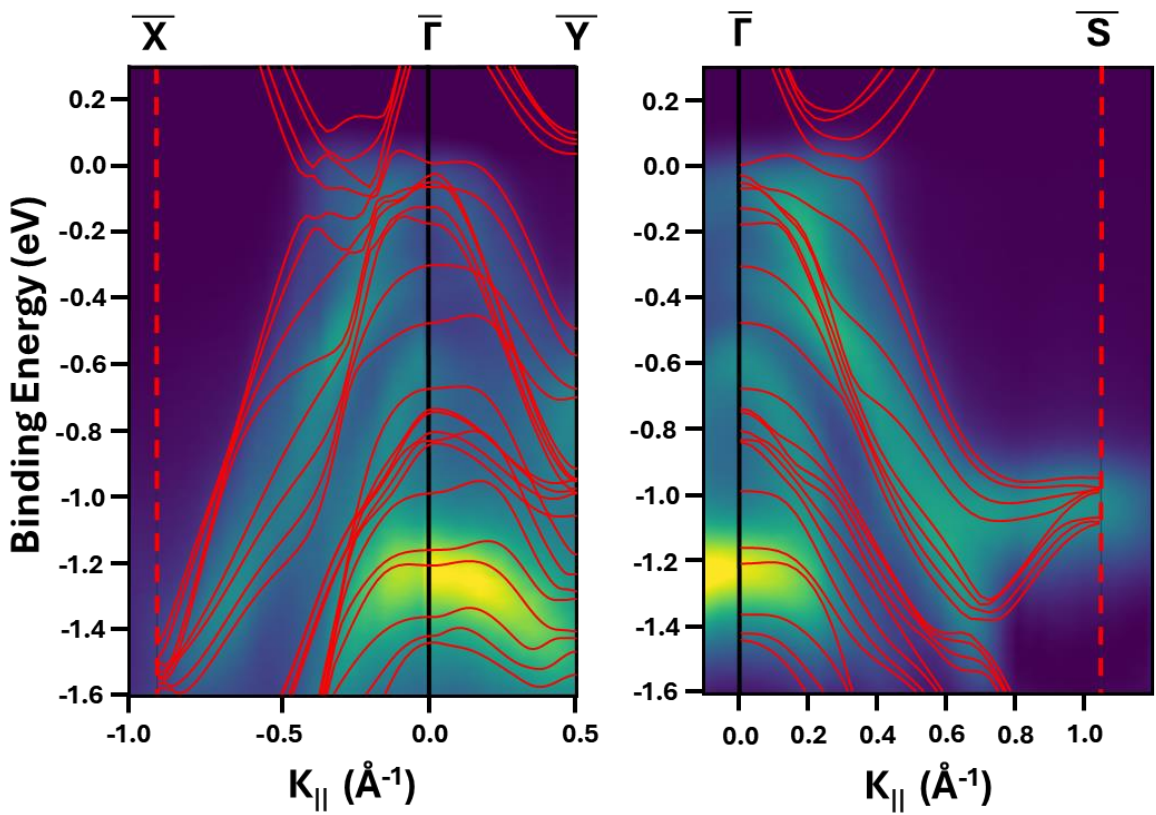

Figure S3. Comparison between the optimized slab calculation band structure and the experimental band structure of pristine 1T'-MoTe$_2$. Left: $\overline{\Gamma X}$ and $\overline{\Gamma Y}$. Right: $\overline{\Gamma S}$.

**Figure S4. DFT Optimized Crystal and Band Structures Using Slab Calculations for Intercalation and Surface Adsorbate Scenarios**

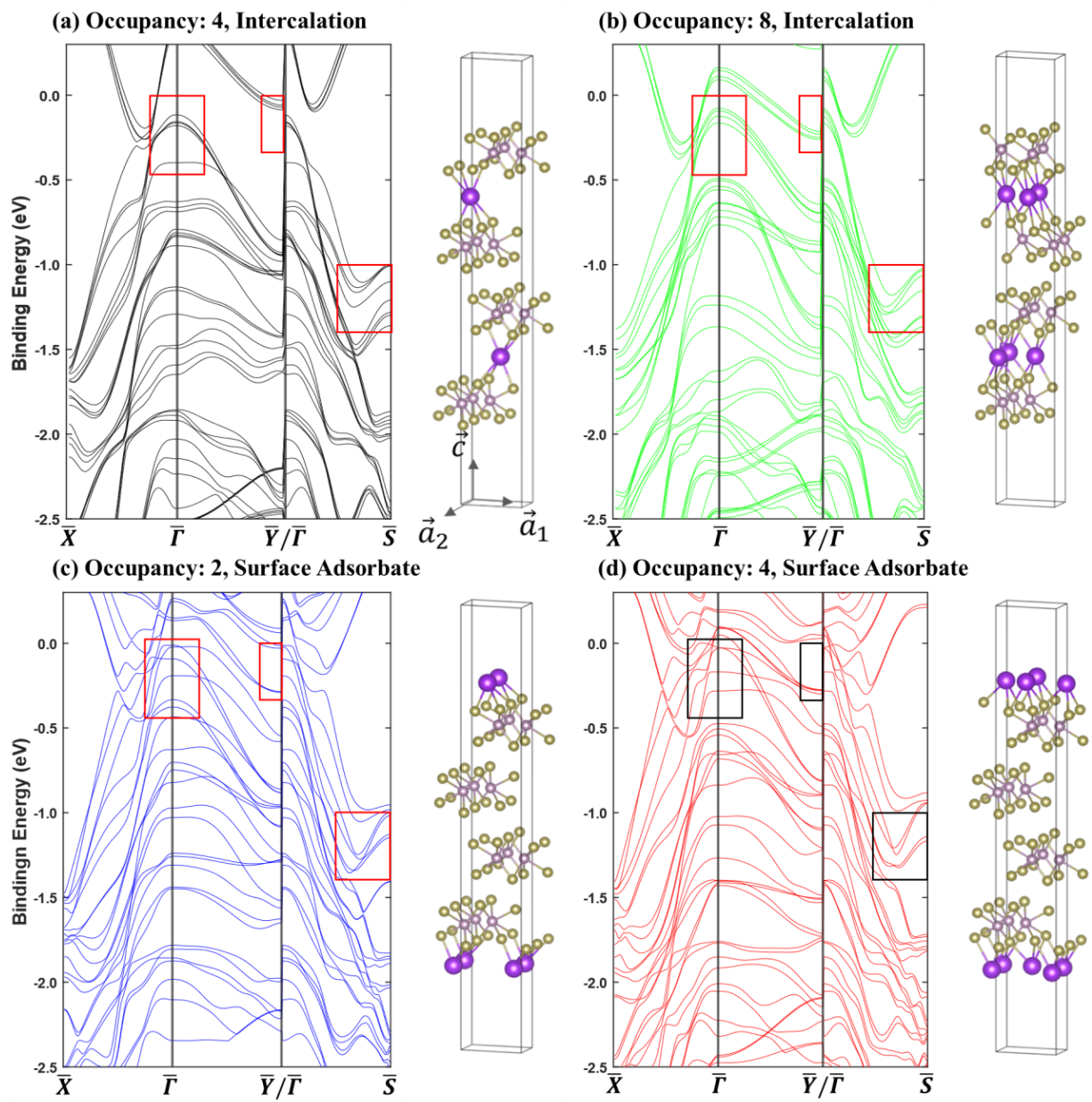

Figure S4. DFT slab calculations of crystal and band structures for select potassium occupation numbers. A discontinuity at $\bar{Y}/\bar{\Gamma}$ occurs because of a jump in the Brillouin zone. (a) The preferred intercalation scenario with an occupancy number of four. Other cases: (b) Intercalation scenario with an occupancy number of eight, the maximum potassium occupation number. (c) Surface adsorbate scenario with an occupancy number of two. (d) Surface adsorbate scenario with an occupancy number of four (maximum number for surface adsorbate). Red boxes (black boxes in S4 (d)) indicate the region where marked differences appear. Crystal structures show the DFT associated optimized slab geometry. Deep purple: potassium, light gold: tellurium, and light purple: molybdenum.

We use a slab calculation method to investigate the band structure of 1T'-MoTe$_2$ with different potassium occupancy numbers and occupation configurations, namely intercalation and surface adsorbates, to assess the most likely experimental scenario. The results presented in Figure S3 are optimized with the procedure mentioned in the main text. Here we show a selection of scenarios that can be ruled out together with the preferred case of occupancy number of four (half-filling, 0.5 K / Mo). An intercalation occupancy number of 8 (full filling, 1.0 K / Mo) supports an excessively pronounced electron pocket at $Y$ (Figure S3 (b)), as well as the absence of a flat band near -0.4 eV at $\Gamma$. The surface adsorbate scenarios can also be readily rejected: The absence of an electron pocket at $Y$ for the full occupancy (full filling, 1.0 K / Mo) in Figure S3 (d) does not match with the ARPES data, and many bands at $\Gamma$ do not fit well with the experimental ARPES data. In the case of half-filling of the surface sites (occupancy number of 2, 0.5 K / Mo) in Figure S3 (c), many bands at the high symmetry points $\Gamma$, $Y$, and $S$ are missing in the experimental data.

# Figure S5. Impact of Higher Potassiation

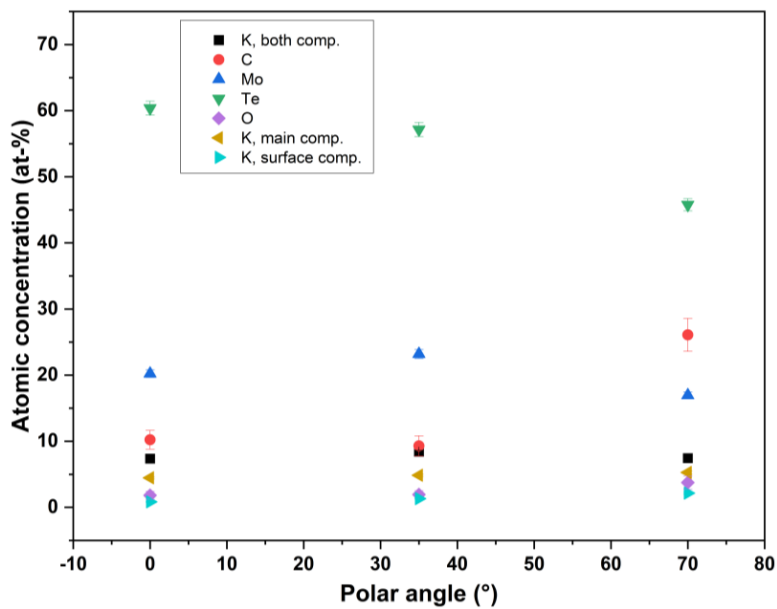

Figure S5. AD-XPS of overdosed K-1T'-MoTe$_2$, showing the atomic concentration of detected elements over select polar angles for a sample with double the potassium load of Figure 5 (b) in the main text, and indicating that excess K diffuses into the crystal and outside the XPS probing depth.

Over the angles we investigated, the K / Mo ratio determined from the values in Figure S4 is nearly identical to those shown in Figure 5 (b) in the main text. It should be noted that the amount of potassium deposited in this sample is twice that in Figure 5 (b). This suggests that thermodynamic equilibrium forces extra potassium atoms to diffuse into deeper layers.

# Figure S6. STM Topography Image of Pristine 1T'-MoTe$_2$ at Room Temperature

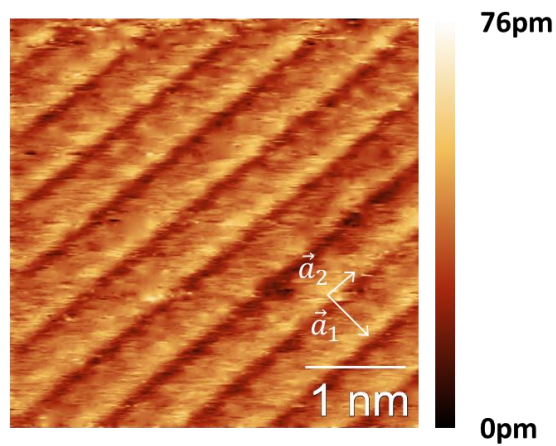

Figure S6. STM image of pristine 1T'-MoTe$_2$ at room temperature with a bias voltage of +150 mV and a current setpoint of 1 nA. Unit cell vector labels are identical to the ones in the main text.

# Figure S7. STM Topography Image of Heavily Potassiated 1T'-MoTe$_2$ at Room Temperature

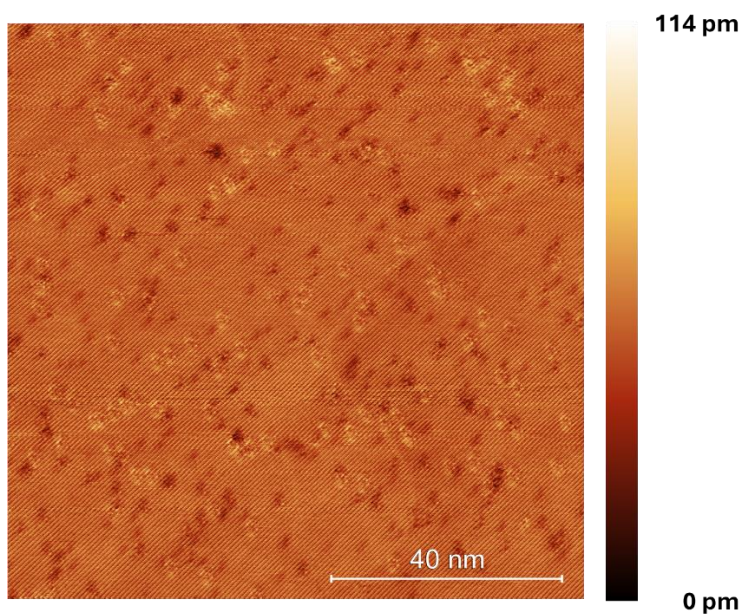

Figure S7. Large scale (100 nm × 100 nm) topography of the heavily potassiated sample. When counting the defect density, the defects are counted based on apparent height variation. Bright protrusions with dark centers are considered to be one defect, not two. The 2D defect density amounts to 0.00827 per MoTe$_2$ unit cell, indicating one defect per 120 unit cells.